\documentclass[12pt]{article}
\begin{document}

\newcommand{\beq}{\begin{equation}}
\newcommand{\eeq}[1]{\label{#1}\end{equation}}
\newcommand{\bea}{\begin{eqnarray}}
\newcommand{\eea}[1]{\label{#1}\end{eqnarray}}
\newcommand{\Tr}{\mbox{Tr}\,}
\newcommand{\bm}[1]{\mbox{\boldmath{$#1$}}}

\begin{titlepage}
\hfill  
\begin{center}
{\Large \bf Gravitational Backreaction \\ of Matter
Inhomogeneities}

\vspace{20pt}

{\large A. Gruzinov$^a$ , M. Kleban$^a$, M. Porrati$^{a,b}$ and
M. Redi$^a$}

\vspace{12pt}

{$^a$ CCPP\\
Department of Physics\\ New York University\\
4 Washington Pl.\\ New York, NY 10003, USA

\vspace{12pt}

$^b$ Scuola Normale Superiore\\ Piazza dei Cavalieri 7\\
I 56126 Pisa (Italy)}
\end{center}

\vspace{20pt}

\begin{abstract}
The non-linearity of Einstein's equations makes it possible
for small-scale matter inhomogeneities to affect the Universe at cosmological
distances. We study the size of such effects using a simple
heuristic model that captures the most important backreaction
effect due to nonrelativistc matter, as well as several exact
solutions describing inhomogeneous and anisotropic expanding
universes. We find that the effects are $O(H^2l^2/c^2)$ or
smaller, where $H$ is the Hubble parameter and $l$ the typical
size scale of inhomogeneities. For virialized structures this is
of order $v^2/c^2$, where $v$ is the characteristic peculiar
velocity.

\end{abstract}

\end{titlepage}

\newpage
\section{Introduction}
The Friedmann-Robertson-Walker (FRW) homogeneous solution to
Einstein's equations gives an excellent approximation to the
large-scale structure of our universe (for a survey, see
e.g.~\cite{kt}). This seemingly uncontroversial assumption is less
obvious than it appears because the distribution of
matter--visible or otherwise--is inhomogeneous already at $\sim
10$ Mpc scales. At that scale the density contrast $\delta \rho /
\rho$ becomes $O(1)$ and the non-linearities of Einstein's
equations makes it conceivable that this may affect some of the
properties of the universe even at cosmological scales.

This possibility has been examined in many papers. Extreme effects
were advocated in~\cite{kmr}, which argued that primordial CMB
inhomogeneities coupled with the non-linear nature of Einstein's
equations could account for the late-time acceleration of the
universe without the need for dark energy. Convincing arguments
against this were advanced in~\cite{hs,iw}, yet the less extreme
claim that inhomogeneity can affect the very large-scale behavior
of the universe is worth investigating. Indeed, even if these
effects turn out to be small, they may still be relevant for
interpreting data of the next generation of cosmological probes.
For instance the study of Ref.~\cite{s} suggests that the
backreaction effect of inhomogeneities at the Hubble scale $H$ is
of the order of $10^{-5}$.

When studying the effect of inhomogeneities, one must be careful in
not mistaking the onset of large {\em matter} inhomogeneities with
the breakdown of the linear approximation for the gravitational
field itself. Intuitively, the difference is that gravity is so weak
that a linear equation for the gravitational field holds everywhere
outside black holes, so, in particular even when $\delta \rho /
\rho\gg 1$ (as on the surface of the Earth). Yet making this
intuition into a precise statement is difficult. One would need a
systematic expansion that takes into account matter non-linearities
to all orders, but treats the gravitational backreaction
perturbatively. That is, one would need an appropriate perturbative
series in $G\delta \rho$~\footnote{$G$ is the Newton constant and
everywhere in this paper $c=1$.}.

We will study this problem heuristically in section 2, by
presenting a ``paradox'' that arises already at second order in
the $G\delta\rho$ expansion, together with its resolution. Namely,
we shall consider point-like particles of mass $m$, distributed on
a regular lattice of size $l$. At scales larger than $l$, one
would expect to find a uniform FRW solution with Hubble constant
$H^2=8\pi Gm/3l^3$. Yet as we will see the backreaction of the
point-like sources is formally infinite. It becomes finite and
 of order $l^2H^2$ only after an appropriate renormalization of the mass $m$.
Physically, this renormalization arises because one cannot separate a
``bare'' mass from its gravitational energy.

Section 2 is heuristic in that the only backreaction term kept
there is the Newtonian gravitational energy. This is not a
rigorous procedure, because effects of similar size are ignored.
To better study backreaction effects we proceed in sections 3 and
4 to study simple exactly soluble cosmologies: an array of equally
spaced parallel two-dimensional dust walls in section 3, and in
section 4 solutions for arbitrary arrangements of parallel cosmic
strings and the ``Swiss Cheese'' model of ref.~\cite{schucking}. 
The symmetries of these toy
models allow us to find explicit solutions of the full Einstein's
equations that again only show effects of $O(l^2H^2)$, consistent
with our general picture.

The toy models will also allow us to address another problem,
namely that whenever inhomogeneities are present, the very
definition of {\em average} cosmological parameters becomes
ambiguous and deserves a thorough re-examination. In the
literature on backreaction effects, the most commonly used
prescription for averages is spatial averages over surfaces of
constant proper time~\cite{b}. This definition is clearly not
physical. The synchronous gauge may not even exist in a general
cosmological solution, and even  when it does it becomes singular
whenever caustics in the matter flow develop. Moreover, it
requires averaging over regions outside our past light cone.
Finally, it gives rise to pitfalls clearly and succinctly
described in~\cite{iw}.

In the toy model of Section 3 we will define several observable
quantities that can be identified with a physical Hubble
parameter. Significantly, these definitions differ from each other
and from the homogeneous result exactly by terms $O(l^2H^2$).

In Section 4 we will find arbitarily inhomogeneous cosmologies for
which the Hubble flow receives no corrections at all, and briefly
discuss the Swiss Cheese model, another inhomogeneous and
anisotropic cosmology in which the local Hubble flow is
uncorrected.

Tellingly, the scale we firmly associate to the large scale effect
of inhomogeneities is $O(v^2)$, with $v$ the typical peculiar
velocity. This result is both physically sensible and large enough
to risk to become a factor in future precision cosmology.

Some technical material omitted in the body of the paper is
collected in two appendices.

\section{Backreaction and Mass Renormalization}

Consider an FRW universe with metric
\beq
ds^2=dt^2 - a^2(t)\gamma_{ij}dx^idx^j, \qquad i,j=1,..,3.
\eeq{m1}
In this section we will often choose a flat homogeneous space metric
$\gamma_{ij}=\delta_{ij}$ to make our (heuristic) equations simpler, but
our results could be easily extended to other homogeneous metrics.

Linearized perturbations of this FRW universe reduce to the
well-known Newtonian limit whenever the stress energy tensor
$T_{mn}$ can be decomposed into a homogeneous piece,
$\bar{T}_0^0=\bar{\rho}$, $\bar{T}_i^j=\delta_i^jp$, sourcing the
Friedmann's equations,  plus an  inhomogeneous ``dust'' component
$t_0^0(\vec{x},t)=\delta\rho(\vec{x},t)$, $t_i^j\approx
0$~\cite{iw}. Here an over-bar will denote space-averaged quantities
(to simplify the discussion we will for the moment ignore
ambiguities in the averaging proceedure, which will be addressed
below). To first order in $G\delta\rho$, the metric is
\beq
ds^2=(1+2\phi)dt^2 - a^2(t)(1-2\phi)\gamma_{ij}dx^idx^j,
\eeq{m2}
while $\phi$ obeys the Poisson equation~\cite{iw}
\beq
a^{-2}D_iD^i\phi =  4\pi G  \delta\rho.
\eeq{m3}
Here $D_i$ is the
covariant derivative w.r.t. the metric $\gamma_{ij}$, which is also
used to raise and lower the indices $i,j$. The dominant non-linear
correction to this equation has a simple physical meaning: gravity
couples to all forms of energy, including the energy of the
gravitational field itself. Thus, to second order in $\phi$,
equation~(\ref{m3}) becomes
\beq
a^{-2}D_iD^i \phi=4\pi G \delta\rho
- {1\over 2} a^{-2} D_i \phi D^i \phi.
\eeq{m4}
Though this equation
is heuristic, it does capture the main effect of non-linearities
since a more complete derivation of backreaction effects leads to a
similar formula~\cite{s,mmb}.

Consider now the case that the background space metric is flat
$\gamma_{ij}=\delta_{ij}$, and that the ``dust'' making up $\delta{\rho}$
is composed of very compact objects of radius $r$
distributed on a cubic lattice of physical
size $l$; by  compact we mean that $r \ll l$. Then
\beq
\delta\rho(\vec{x})=
\sum_{\vec{n}\in Z^3} \rho_r\left(a\vec{x}-l \vec{n}\right)-\bar{\rho} .
\eeq{m5}
Here $\rho_r(\vec{x})$ is an arbitrary positive function vanishing for
$|\vec{x}|>r$ and normalized to
\beq
\int d^3 x \rho_r(\vec{x})=m.
\eeq{m6}

The backreaction equation~(\ref{m4}) does not contain any time 
derivative, so, when studying
the effect of inhomogeneities at any given time, as we will do next,
we can set the scale factor $a=1$ to avoid
needless complications.

The average background density $\bar{\rho}$ is a function of the
cluster mass $m$ and the volume of the fundamental cell of the lattice
$\bar{\rho}= m/ l^3.$

To understand the large scale effect of the non-linear term we average
eq.~(\ref{m4}) over a cube of side much larger than $l$ at constant time $t$.
The averaging procedure itself introduces an ambiguity, since we could have
chosen to average over a different space-like surface. In this section, this
ambiguity will not matter, since here we merely want to show how to avoid
much larger, indeed divergent, unphysical effects. 

The average $\overline{\delta{\rho}}$ is zero by construction,
so eq.~(\ref{m4}) averages to
\bea
\triangle \bar{\phi}&=& -{1\over 2}\overline{\nabla \phi \nabla \phi},
\label{m8}\\
\overline{\nabla \phi \nabla \phi}&=&\lim_{V\rightarrow \infty} {1\over V}
\int_V d^3x \nabla \phi \nabla \phi.
\eea{m9}
The gradient and the Laplacian are the standard flat-space ones.
To second order in $G\delta\rho$, the potential $\phi$
appearing in the right hand side of
this equation is the solution of the Poisson eq.~(\ref{m4}).

A brief computation then shows that the average gravitational energy density is

\beq
-{1\over 8\pi G}\overline{\nabla \phi \nabla \phi} =
-{ G\over l^6} \sum_{\vec{m}\in Z^3,\vec{m}\neq 0}{l^2 \over 2\pi m^2}
\left| \tilde{\rho}_r \left({2\pi\over l}\vec{m}\right)\right|^2,
\eeq{m11}
where 
$\tilde{\rho}_r(\vec{k})\equiv \int d^3 x 
\rho_r(\vec{x})\exp(i\vec{k}\vec{x})$.

Here we encounter a (pseudo) paradox: when the ``clusters'' are
point-like, the distribution $\tilde{\rho}_r$ is
a constant and sum in eq.~(\ref{m11}) diverges.
If we regulate the divergence--say by
making $\rho_r(\vec{x})= 3/4\pi r^3$ for $|\vec{x}| \leq r$ and zero
otherwise--we may still end up with a gravitational energy density
as large as the background energy $\bar{T}_0^0$. This happens in particular if
the radius of the sphere is close to its Schwarzschild
radius. Physically, this would mean that if all matter in a
universe at critical density was distributed into black holes, corrections to
the linear approximation would be $O(1)$ {\em no matter how small the lattice
step $b$}!.
Even when $r\gg Gm$ this contribution is suspicious because it
depends strongly on the size of the cluster.

An estimate of the  $r$ dependence of the sum is given in the Appendix, here we only quote
the result, and we give a simple argument to justify it:
\beq
-{1\over 8\pi G}\overline{\nabla \phi \nabla \phi} =
-Gm^2\left[ {C_1\over r l^3}+{C_2\over l^4} +O\left({r\over l^5}\right) ~ \right],
\eeq{m14}
where $C_1$, $C_2$ are dimensionless numbers of order unity which are determined by the
density profile of compact objects $\rho _r$ and by the shape of the
lattice. The first term in the expansion (\ref{m14}) comes from the
classical gravitational energy of a body of mass $m$ and size $r$.
Other terms, describing corrections, must be there because for a
body of uniform density filling the entire lattice cell
$\bar{t}_0^0$ vanishes by definition. The divergence at small $r$
term scales with $l$ as pressureless mass density, while the finite
term, independent of $r$, scales as ultra-relativistic energy
density.

At this point, it is obvious that the divergent term is unphysical
and it must be canceled by redefining the ``bare'' mass of the
compact objects to first order in $G$ by
\beq m=m_{physical} +
C_1{Gm^2\over r }.
\eeq{m15}
This {\em classical} renormalization is
{\em not} an option because the gravitational field is actually
determined by the physical mass.
\beq
m -{G\over 2}\int d^3 x \rho_r(0)  {1\over | \vec{x}|} \rho_r(\vec{x})=
m - G\int {d^3k\over (2\pi)^3} {2\pi\over k^2}
\left| \tilde{\rho}_r (\vec{k})\right|^2\equiv m_{physical}.
\eeq{m16}.

Notice that corrections due to inhomogeneities do exist. They are
due to the finite term in eq.~(\ref{m14}), which cannot be
eliminated by a renormalization of the mass, since it scales with
$l$ as radiation. These corrections change the Friedmann equation by
terms $O(H^2l^2)$. If $l$ is interpreted as the non-linear length
scale of the actual universe, then $H^2l^2\sim v^2$, where $v$ is
the typical peculiar velocity. So, non-linear corrections to the
Hubble parameter due to gravitational energy are of the same order
as relativistic corrections due  to peculiar velocities. We have
actually neglected the latter as well as ambiguities in the
definition of the Hubble parameter. Evidently, to make further
progress we need to be more systematic, even if at the price of
studying a vastly simplified model of inhomogeneities. This is what
we will do next, starting by considering a distribution of
pressureless matter that only breaks translational invariance in one
direction, while preserving translations and rotation in two
orthogonal directions.

\section{Dust-Wall Universe}

Our goal is to construct an exact solution of the Einstein's
equations describing an expanding inhomogeneous universe. A simple
possibility is planar symmetry -- the metric depending on $t$ and
$x$ but not on $y$ and $z$, and isotropic in the $yz$ plane. Here
we use Taub's \cite{taub} explicit expressions for the metric of
vacuum plane-symmetric spacetimes to construct a universe of
equidistant plane-parallel dust walls.  Our treatment of the walls
is similar to the single wall case \cite{vilenkin,ipser}.

The plane-symmetric metric can be written as
\begin{equation}
ds^2=e^{2u}(dt^2-dx^2)-e^{2v}(dy^2+dz^2), \label{a1}
\end{equation}
where $u=u(t,x)$ and $v=v(t,x)$. The walls are located at $x=\pm
b,~\pm 3b, ...$ so that the metric coefficients $u$ and $v$ are
$2b$-periodic functions of $x$. It also follows from the symmetry of
the problem that $u$ and $v$ are even functions of $x$.

The stress-energy tensor of a thin dust wall of proper surface
density $\sigma (t)$ located at $x=b$ is
\begin{equation}
T^\mu_\nu(t,x)=\sigma (t) e^{-u(t,b)}\delta (x-b) {\,\rm
diag}(1,0,0,0). \label{a2}
\end{equation}
For simplicity we consider walls with zero pressure, but it is
easy to find the analogous solutions for an array of walls with
general equation of state. Calculating the Einstein tensor
$G^\mu_\nu$ for the metric (\ref{a1}), equating it to the
stress-energy tensor of the dust walls, and using the symmetry of
the metric, we get the following jump conditions at the $x=b$
wall: the $G^0_0$-component gives
\begin{equation}
\sigma=4e^{-u}\partial _xv, \label{a3}
\end{equation}
the $G^0_1$ and $G^1_1$ components are non-singular at the wall, and the
$G^2_2=G^3_3$ components give
\begin{equation}
\partial _x(u+v)=0,
\label{a4}
\end{equation}
where the derivatives are calculated for $x=b-0$.

Between the walls the metric is given by the Taub's expressions:
\begin{equation}
e^{2u}={f'g'\over \sqrt{f+g}},~~~e^{2v}=f+g,~~~f=f(t+x),~~~g=g(t-x),
\label{taub}
\end{equation}
where $f$ and $g$ are arbitrary functions of one variable, and $f'$
and $g'$ are their derivatives. Since $u$ and $v$ are even functions
of $x$ at all time, we must take $g(\xi )=f(\xi )$. Then (\ref{a4})
gives an equation for $f$:
\begin{equation}
{f''(t+b)\over f'(t+b)}-{f''(t-b)\over f'(t-b)}+{1\over
2}{f'(t+b)-f'(t-b)\over f(t+b)+f(t-b)}=0. \label{a5}
\end{equation}
The remaining jump condition (\ref{a3}) gives the proper surface
density as a function of time in terms of $f$.

At large times $t\gg b$ there are many walls within the horizon and
we recover the homogeneous matter-dominated universe. To lowest
order in $b$, finite differences in equation (\ref{a5}) can be
replaced by differentials, and we get
\begin{equation}
\left( {f''\over f'}\right) '+{1\over 4}{f''\over f}=0, \label{a6}
\end{equation}
with the solution $f(t)\propto t^4$. To lowest order in $b$ this
gives the metric $e^{2u}\propto e^{2v}\propto t^4$. This describes
homogeneous matter-dominated universe with conformal time $t$.

As expected, there are finite $b$ corrections. In the next to
leading order, we get $f(t)\propto t^4+40b^2t^2$. This gives the
metric coefficients
\begin{equation}
e^{2u}\propto t^4+t^2(20b^2-6x^2), ~~~ e^{2v}\propto
t^4+t^2(40b^2+6x^2). \label{a7}
\end{equation}
We see that the metric is close to that of a uniform matter
dominated universe, with $b^2/t^2$ corrections. In physical units
the corrections are $\sim H^2l^2$, where $l$ is the physical
distance between the walls and $H$ is the Hubble constant.

Now we would like to see how the averaged expansion rate of the
universe is changed due to inhomogeneities. To this end, we define
parallel and perpendicular scale factors on the wall:
\begin{equation}
a_{\parallel }(t)\propto e^{v(t,b)}, ~~~a_{\perp }(t)\propto
e^{u(t,b)}. \label{a8}
\end{equation}
In the next to leading order we get
\begin{equation}
a_{\parallel }(t)\propto t^2+23b^2, ~~~a_{\perp }(t)\propto
t^2+7b^2. \label{a9}
\end{equation}
Further we define parallel and perpendicular Hubble constants
\begin{equation}
H_{\parallel }={d\log a_{\parallel }\over d\tau }, ~~~H_{\perp
}={d\log a_{\perp }\over d\tau }, \label{A10}
\end{equation}
where $d\tau \propto a_{\perp }dt$ is the proper time on the wall.
Both definitions of the Hubble constants are physical. $H_{\parallel }$
can be measured from the proper surface energy density as a function
of proper time. $H_{\perp }$ can be measured by light propagation
time between the walls. We then get corrections to the Friedmann
equation for parallel and perpendicular Hubble constants
\begin{equation}
H_{\parallel }^2a_{\parallel }^3\propto 1+{9\over 16}H^2l^2, ~~~
H_{\perp }^2a_{\perp }^3\propto 1-{7\over 16}H^2l^2, \label{a11}
\end{equation}
where $l\approx 2be^u$ is the physical distance between the walls and $H\approx H_{\parallel }\approx H_{\perp }\approx 2/(3\tau )$.

\section{Uncorrected Hubble Flow}
\label{2+1}

In this section we will consider a broad class of examples of
anisotropic cosmologies for which we can find exact solutions,
this time sourced by arrangements of parallel relativistic cosmic
strings rather than parallel walls. The dynamics of such
objects--so long as they remain parallel--is identical to that of
massive point particles in 2+1 dimensional gravity, and so for the
rest of the section we will use that language.  We will show that
starting with \emph{any} isotropic and homogenous solution of
Einstein's equations in 2+1 dimensions one can add point
particles--or indeed an arbitrary distribution of dust--and 
(due to the special properties of gravity in
2+1 dimensions) the presence of the point particles and/or dust
does not modify the Hubble expansion at all.

We begin with the metric
\begin{equation}
ds^2 = dt^2 - a(t)^2 ds_2^2, \label{mm1}
\end{equation}
where as usual $ds_2^2$ is the metric of a homogeneous and
isotropic two-dimensional flat, spherical, or hyperbolic space.
From this ans\"{a}tz the 2+1 Friedmann equation follows:
\begin{equation}
H^2 = \rho - k/a^2, \label{mm2}
\end{equation}
where as usual $k= \{ -1,0,+1 \}$, $H = \dot a/a$ and $\rho =
\rho_0(a_0/a)^{2(1+w)}$ for the case of a perfect fluid with
equation of state $w$.\footnote{$w$ refers to the 2+1 dimensional
equation of state. Lifting $w=0$ here to 3+1 gives an anisotropic
but homogeneous 3+1 dimensional universe filled with a uniform
``gas'' of parallel cosmic strings.}

One peculiar feature special to the case of matter domination
($w=0$) is that $a(t) = t$, with zero acceleration. This indicates
that uniformly expanding dust in 2+1 dimensions does not exert a
force on itself, a result which might have been expected since there
is no force between static particles in 2+1 dimensions. Note also
that, since in this case the matter density scales the same way as
curvature, the scale factor is the same for all three types of
universes.

\subsection{Point Particles}

In order to consider an anisotropic cosmology we begin by
reviewing the physics of a point particle with mass $m \in
(0,2\pi)$ in a 2+1 dimensional space\footnote{We will work in
units where $8 \pi G_{3} = 1.$}. As is well-known the metric is
simply flat space with a conical defect along the world-line of
the particle: $ds_2^2 = dt^2 - dr^2 - r^2 d\theta^2$, where the
range of $\theta$ is from $0$ to $2 \pi-m$. Perhaps the simplest
way to see this is to recall that in 2+1 dimensions Einstein's
equation in vacuum $R_{\mu \nu} = 0$ implies $R_{\mu \nu \lambda
\sigma} = 0$, so that the space must be flat away from the point
particle. The conical defect induces a delta function in the
curvature of strength $2 m$, from which it follows that the metric
satisfies Einstein's equations with a point-like source. Since any
space is locally flat, all of the solutions we consider here are
of this form very close to the singularity. Note that the mass of
the particle should be smaller than $2\pi$ since the circumference
of the circle goes to zero in that limit.

Let us consider a collection of point particles with masses $m_i$ at
locations $z_i$ in flat space. Taking the metric ans\"{a}tz
\begin{equation}
ds^2=dt^2 - e^{\phi(z,\bar{z})} dz d\bar{z}, \label{mm3}
\end{equation}
Einstein's equations reduce to
\begin{equation}
\partial_z\partial_{\bar{z}}\, \phi= - \sum_{i=1}^N m_i
\delta^2(z-z_i). \label{mm4}
\end{equation}
So long as the sum of the masses satisfies the constraint $M =
\sum_{i=1}^N m_i < 2 \pi$ the solution is
\begin{equation}
e^{\phi}= \displaystyle{\Pi}_{i=1}^N |z-z_i|^{-\frac {m_i} \pi}.
\label{mm5}
\end{equation}
It is easy to see where the constraint on the total mass comes from.
Consider a large disk $S$ containing all the masses. Recall that the
Euler character $\chi=1$ for a disk, that the space is flat away
from the point particles, and that a point particle of mass $m$
gives rise to a delta function in the scalar curvature $R$ of
strength $2m$. Then, since the integrated extrinsic curvature of a
circle is the derivative of its circumference with respect to the
radius, $\partial_r c$, the Gauss-Bonnet theorem,
\begin{equation}
\int_S R + 2 \int_{\partial S} K = 4 \pi \chi, \label{mm6}
\end{equation}
implies $\sum m_i = 2 \pi - \partial_r c$. If the total mass $M$
exceeds $2 \pi$ the space cannot be globally flat, since the
circumference of the disk must shrink as the radius increases. If we
insist that the curvature be zero, the solution (\ref{mm5}) now
implies the presence of a conical singularity at $z=\infty$ equal to
$4\pi - \sum_{i=1}^N m_i$. The metric then describes a compact space
with spherical topology, with $N+1$ conical singularities whose
total mass adds up to $4 \pi$.

From the Gauss-Bonnet theorem it follows that in order to
accommodate point particles with total mass $M > 4 \pi$ some
curvature is needed. We wish to consider the case of a
large or infinite number of (positive) masses, and so the
background 2-dimensional curvature must be negative. 
In fact the expansion of the universe automatically acts as a source
for the negative spatial curvature.

Consider a metric of the form
\begin{equation}
\label{mm7} ds^2 = dt^2 - a(t)^2 e^{\phi(z,\bar{z})}dz d\bar{z}.
\end{equation}
If we now require that the energy-momentum tensor consists of
point particles of mass $m_i$ at fixed spatial locations $z_i$
plus a homogeneous background matter density $\rho_{\rm hom}$,
Einstein's equations reduce to
\begin{eqnarray}
\partial_z\partial_{\bar{z}}\, \phi &=& - \frac k 2\, e^\phi - \sum_{i=1}^N
m_i\,
\delta^2(z-z_i),\nonumber \\
\left( {\dot a \over a} \right)^2 &=& \rho_{\rm hom}-\frac k
{a^2}, \label{mm8}
\end{eqnarray}
where $k$ determines the (constant) spatial curvature away from
the singularities.  This is the main result of this section. As is clear from the
second equation the evolution of the scale factor is not disturbed
by the defects as long as the first equation can be satisfied.
This in turn depends only on the masses $m_i$ (see appendix). The
conclusions above hold even in the presence of general homogenous
matter components. In this case the pressure of the homogenous background
is given by, 
\begin{equation}
p=-\frac {\ddot{a}} a.
\end{equation}

For the case $k=0$ the first equation reduces to the Poisson
equation, whose solution with point particle masses we discussed
above. When $k=\pm 1$ eq. (\ref{mm8}) is the Liouville equation with
positive and negative curvature respectively. As we discuss in the
appendix, solutions to this equation always exist, unless
topological obstructions arise. From the Liouville equation it also
follows that, as required by the Gauss-Bonnet theorem, the local
density of particles cannot be too large to avoid ``overclosing''
the universe.

One case of interest is a square grid with a point mass at each
vertex and with the background matter density $\rho_{\rm hom} =0$.
Since the total mass in the universe is greater than $4 \pi$ the
background curvature must be negative. The mathematical problem we
need to solve reduces to finding the metric with constant negative
curvature and a single conical singularity on a square torus.
Although in this case the Liouville equation cannot be solved in
closed form, the solution is known to exist and to be unique.
Physically, while the metric has locally negative spatial
curvature away from the particles, it has precisely the correct
density of positively curved defects to cancel the negative
curvature when averaged over a large region (or over one tile of
the grid). The solution approximates flat FRW at large scales (and
it is for this reason that it is possible to arrange the masses in
a square grid even though the metric locally is negatively
curved).

In general there is a very large continuous infinity of solutions
of this type (corresponding to the freedom of changing the
positions and masses of the particles). In fact (as is clear since
we can take the limite of very small strength $\delta$-functions
closely spaced) if we replace the term $\sum_{i=1}^N m_i\,
\delta^2(z-z_i)$ in equation (\ref{mm8}) with a function
$\rho_{\rm dust}(z, \bar z)$ we can obtain solutions with
arbitrary distributions of dust.

It is remarkable that the function $a(t)$ in the metric
(\ref{mm1}) remains unchanged for \emph{any} arrangement of point
masses or distribution of dust, and that therefore the local Hubble
expansion law is entirely uncorrected by the inhomogeneities. This
is a physical statement confirmable by local experiments; for
example an observer in this universe could measure the rate of
change in the local matter density as a function of his proper
time and would observe it to be determined only by the average
Hubble constant (averages here can be taken over surfaces of
constant $t$, for which the expansion is uniform), independent of
his location and the locations of the anisotropies.

Let us mention that, even though the Hubble expansion is
not modified, many experiments will be sensitive to the
positions of the masses, for example test particles fired out
along some trajectory will scatter off the point masses.  For the
same reason the observed brightness of a distribution of standard
candles of fixed red-shift will vary.

\subsection{Swiss Cheese}
There exists in the literature another exact solution describing
an inhomogeneous universe which gives no corrections to the Hubble
parameter--the ``Swiss Cheese" model of Schucking
\cite{schucking}. This is a universe in which spherical regions of
homogeneous background matter are excised and replaced with
vacuum, but with a spherical concentration of mass at the center
of the void.  If the mass at the center is chosen appropriately
and the voids are non-intersecting the metric outside the voids is
completely unaffected by their existence and the Hubble flow as
measured by an (outside) observer is unchanged.  Of course as
before there are many experiments which would be sensitive to the
inhomogeneities.

\section{Conclusion}
Both generic (but heuristic) as well as exact (but highly
symmetric) models show that the effect of matter inhomogeneities
on the large scale expansion of the universe is small, bounded
from above by $v^2\sim 10^{-5}$.  The numerical coefficient
and sign of the effect depend on the model and the definition
chosen for Hubble. What we have shown can also be interpreted as
follows: there always exist gauges where the metric differs
from the FRW metric only at order $\sim v^2$.

\subsection*{Acknowledgments}
We thank P. Creminelli, B. Menard, A. Nicolis, N. Kaloper,  R.
Scoccimarro and E. Sheldon for useful discussions. We especially
thank E. Schucking for the same and for pointing out to us
ref.~\cite{schucking}. A.G. is supported by the David and Lucile
Packard foundation, M.P. is supported in part by NSF grant
PHY-0245068, and by a Marie Curie chair, contract
MEXC-CT-2003-509748 (SAG@SNS). M.R. is supported by the NSF grant
PHY-0245068.

\section*{Appendix A: Estimating the Gravitational Self-Energy Divergence}
\renewcommand{\theequation}{A.\arabic{equation}}
\setcounter{equation}{0}
Here we estimate the behavior of the sum in eq.~(\ref{m11}).

The solution of the Poisson eq.~(\ref{m4}) is easily written in
Fourier transform  as
\beq
\tilde{\phi}(\vec{k})= -{4\pi G \over k^2} \sum_{\vec{m}\in Z^3,\vec{m}\neq 0}
{(2\pi)^3\over l^3} \delta^3\left(\vec{k}- {2\pi\over l} \vec{m}\right)
\tilde{\rho}_r(\vec{k}).
\eeq{m10}

For $l\rightarrow \infty$, eq.~(\ref{m11}) can be approximated by the integral
\beq
\lim_{l\rightarrow \infty}{1\over l^3}\sum_{\vec{m}\in Z^3,\vec{m}\neq 0}
{l^2 \over 2\pi m^2}
\left| \tilde{\rho}_r \left({2\pi\over l}\vec{m}\right)\right|^2
= \int {d^3k\over (2\pi)^3} {2\pi\over k^2}
\left| \tilde{\rho}_r (\vec{k})\right|^2.
\eeq{m12}
Since $\tilde{\rho}_r(\vec{k})$ is analytic and square summable, we can also
estimate the sum at finite $l$ as
\beq
{1\over l^3}\sum_{\vec{m}\in Z^3,\vec{m}\neq 0}{l^2 \over 2\pi m^2}
\left| \tilde{\rho}_r \left({2\pi\over l}\vec{m}\right)\right|^2
={m^2\over r} f\left({r\over l} \right).
\eeq{m13}
By dimensional analysis, $f(x)$ is a dimensionless function, independent
of $m,r,l$, and smooth in a neighborhood of $x=0$.

Therefore we can see that the energy density diverges 
as $1/r$ for $r\rightarrow 0$,
and we also recover the finite term of eq.~(\ref{m14}).

\section*{Appendix B: More on the Liouville Equation}
\renewcommand{\theequation}{B.\arabic{equation}}
\setcounter{equation}{0}
As we have shown in section \ref{2+1}, the problem of point
particles (with zero peculiar velocities) in an homogenous FRW
universe in 2+1 dimensions reduces to finding solutions to the
Liouville equation,
\begin{equation}
\partial_z\partial_{\bar{z}}\, \phi=-\frac k 2\,  e^\phi - \sum_{i=1}^N
m_i\, \delta^2(z-z_i). \label{aa1}
\end{equation}
This equation describes a two dimensional surface of constant
curvature $k$ with conical singularities $m_i$. The general solution
of the Liouville equation is
\begin{equation}
e^\phi=\frac {4 |w'|^2}{\left[1+k |w|^2\right]^2},
\label{aa2}
\end{equation}
where $w(z)$ is a holomorphic function that must be chosen
appropriately in order to reproduce the correct singularities at
$z_i$\footnote{This asymptotic behavior holds for $m_i<2\pi$. When
$k$ is negative singularities with $m_i=2\pi$ (so called parabolic)
are also acceptable in which case $\phi\sim -2 \log |z-z_i|-2 \log |\log |z-z_i||$.},
\begin{equation}
\phi \sim - \frac {m_i} {\pi} \log |z-z_i|, ~~~~~~~~~~ \mbox{as}~~
z\to z_i.
\end{equation}
In principle $w(z)$ can be determined by using the technology of the
Fuchsian equations (see for example \cite{takhtajan}), even though
explicit solutions can only be found in special cases. The solutions
of the Liouville equation are in general determined by the
singularities and by the topology of the space. Since the locations
$z_i$ of the singularities in eq. (\ref{aa1}) are arbitrary (up to
conformal transformations) there is actually a large moduli space of
solutions. Note that even though the $z_i$'s are free, the position
in physical space are constrained since on scales smaller than the
curvature the total deficit angle cannot exceed $2\pi$. This is
automatically guaranteed by solving the Liouville equation.

Let us now consider the case where the space is compact. By
integrating both sides of eq. (\ref{aa1}) one finds,
\begin{equation}
V=\frac {4\pi(1-g) -\sum_{i=i}^N m_i} k, \label{aa3}
\end{equation}
where $V$ is the volume and $g$ is the genus of the surface. This
formula follows from the fact that the left hand side of eq.
(\ref{aa1}) is proportional to $\sqrt{g}R$ and the Gauss-Bonnet
formula for compact surfaces without boundaries,
\begin{equation}
\frac 1 {4 \pi}\int_S \sqrt{g} R =2 -2 g. \label{aa4}
\end{equation}
Since the volume must be positive, eq. (\ref{aa3}) implies
restrictions on the allowed singularities depending on the sign of
the curvature.

\subsection*{$k=1$}

This situation can arise for example in the presence of background
dust or positive cosmological constant. Since we only allow positive
masses the volume formula (\ref{aa3}) here determines that the only
possible topology is the spherical one and that the sum of the
deficit angles must be less than $4\pi$. The Liouville equation can
be solved explicitly in the case of three singularities. In general
a solution exists provided that a mild constraint on the deficit
angles is satisfied, namely that the largest conical defect is less
than the sum of the remaining ones \cite{luo}.

\subsection*{$k=-1$}

This is the case relevant for particles in vacuum when the sum of
the deficit angles is greater than $4\pi$. An important theorem
dating back to Poincar\'e and Picard establishes that when the
curvature is negative a solution of the Liouville equation with
prescribed singularities always exists and is unique, once 
eq.~(\ref{aa3}) is satisfied. If the sum of the deficit angles is
greater than $4\pi$ any topology is allowed and in particular we can
choose the spherical topology. Note that in hyperbolic space a
deficit angle increases the volume of the space. A case relevant for
our purposes is the torus with one conical singularity. This problem
is equivalent to that of a grid of equal masses. From
eq.~(\ref{aa3}) we derive that the volume of each tile equals the
deficit angle. This ensures that the average curvature is zero as
expected on physical grounds.


\end{document}